\documentclass[12pt]{article}
\usepackage{aaspp}
\lefthead{}
\righthead{Pulsar Wind Nebula in SNR G292.0_1.8}
\def\CXO{{\it Chandra X-ray Observatory}}
\def\chandra{{\it Chandra}}
\def\g292{G292.0+1.8}
\def\msh1154{MSH 11$-$5{\sl 4}}

\def\rosat{{\it ROSAT}}
\def\asca{{\it ASCA}}

\def\chandra{{\it Chandra}}

\def\lsim{\hbox{\raise.35ex\rlap{$<$}\lower.6ex\hbox{$\sim$}\ }}
\def\gsim{\hbox{\raise.35ex\rlap{$>$}\lower.6ex\hbox{$\sim$}\ }}

\begin{document}

\title{A Pulsar Wind Nebula in the Oxygen-Rich Supernova Remnant
\g292}

\author{John P. Hughes\altaffilmark{1},
Patrick O.~Slane \altaffilmark{2},
David N.~Burrows\altaffilmark{3},
Gordon Garmire\altaffilmark{3}, 
and John A.~Nousek\altaffilmark{3}}
\altaffiltext{1}{Department of Physics and Astronomy, Rutgers
University, 136 Frelinghuysen Road, Piscataway, NJ 08854-8109;
jph@physics.rutgers.edu}
\altaffiltext{2}{Harvard-Smithsonian Center for Astrophysics, 60 Garden Street,
Cambridge, MA 02138; slane@head-cfa.harvard.edu}
\altaffiltext{3}{Department of Astronomy \& Astrophysics, 525 Davey Lab,
Penn State University, University Park, PA 16802}  

\begin{abstract}

Using the \CXO\ we have discovered a diffuse, center-filled region of
hard X-ray emission within the young, oxygen-rich supernova remnant
(SNR) \g292. Near the peak of this diffuse emission lies a point-like
source of X-ray emission that is well described by an absorbed
power-law spectrum with photon index $1.72\pm0.09$.  This source
appears to be marginally extended; its extent of 1.3$^{\prime\prime}$
(FWHM) is greater than that of a nearby serendipitous X-ray source
with FWHM = $1.1^{\prime\prime}$.
This is strong evidence for the presence within SNR \g292\ of a young
rapidly-rotating pulsar and its associated pulsar wind nebula.  From
the unabsorbed, 0.2-4 keV band X-ray luminosity of the pulsar wind
nebula ($L_X\sim 4\times 10^{34}\,\rm ergs\,\, s^{-1}$), we infer a
spin-down energy loss rate of $\dot E \sim 7\times 10^{36}\,\rm
ergs\,\, s^{-1}$ for the still undetected pulsar. The pulsar candidate
is $0.9^\prime$ from the geometric center of the SNR which
implies a transverse velocity of $\sim$$770 (D/4.8\,{\rm kpc})
(t/1600\,{\rm yr})^{-1}\,\rm km \,s^{-1}$ assuming currently accepted
values for the distance and age of \g292. 

\end{abstract}


\section{Introduction}

\g292\ (\msh1154) is a southern supernova remnant (SNR), discovered in
the radio band (Milne 1969; Shaver \& Goss 1970), with a radio
spectral index intermediate between that of a typical shell-type
and Crab-type remnant (Goss et al.~1979).  Optically the
spectrum is dominated by oxygen and neon emission lines (Goss et
al.~1979) covering a wide range, $\sim$2000 km s$^{-1}$, in velocity
(Murdin \& Clark 1979).  \g292\ is, therefore, an oxygen-rich SNR, of
which there are only two other examples in the Galaxy (Cassiopeia A
and Puppis A) and a handful elsewhere (e.g., N132D, E0540$-$69.3, and
E0102.2$-$7219 in the Magellanic Clouds).  It is believed that \g292\
is a young SNR with an age $\lsim$1600 yr in which the supernova
ejecta have not yet fully mixed with the swept-up circumstellar
material and thus may represent a slightly older version of Cas A.

Recent developments suggest that \g292\ is, in fact, a composite
remnant, i.e., one showing both thermal emission from SN ejecta, as
well as a synchrotron nebula powered by a rapidly rotating neutron
star.  Wallace \& Gaensler (1997) presented a high resolution 20-cm
radio map of the SNR from the Australia Telescope, which shows a
bright, extended nebula near the center with a flatter spectrum
($\nu^{-0.4}$) than elsewhere in the remnant ($\nu^{-0.7}$).
Pulsar-powered plerions, like the Crab Nebula, have traditionally been
distinguished from their shell-type cousins by their flatter radio
spectra.  Recently Torii, Tsunemi, \& Slane (1998) have reported on
the \g292\ \asca\ data.  In addition to the thermal emission at soft
X-ray energies, above $\sim$3 keV they found a hard X-ray source with
a power-law spectrum, unresolved at \asca\ angular resolution and
positionally consistent (within the usual \asca\ position uncertainty
of 40$^{\prime\prime}$) with the extended, flat-spectrum radio
component mentioned above. Although not conclusive, these observations
suggest that a synchrotron nebula powered by the spin-down of a
central pulsar possibly lurks within \g292.  This finding is quite
important, since it would allow us to conclusively associate the
young, oxygen-rich SNR \g292\ with a core-collapse, massive star SN
explosion.  Indeed only one other oxygen-rich remnant, SNR
E0540$-$69.3 in the Large Magellanic Cloud, is known to contain a
rapidly spinning pulsar.

\section{Observations}

\g292\ was observed with the back-side illuminated chip (S3) of the
\chandra\ Advanced CCD Imaging Spectrometer (ACIS-S) (Garmire et
al.~1992; Bautz et al.~1998) on 11 March, 2000, as part of the Penn
State GTO program. Observations were taken in full frame mode with a
readout time of 3.2 s.  We filtered the data for times of flaring
background or bad aspect solution, applied the bad pixel map, and
verified that the level 2 events had been gain-corrected using the
gain map appropriate to the focal plane temperature at the time of
observation ($-$120 C).  The effective deadtime-corrected exposure
time after time filtering was 43026.6 s.

The S3 chip was used in order to take advantage of its soft response
and good spectral resolution.  The pointing direction was set so that
as much of the remnant at possible fell on this chip. The remnant is
slightly larger than $8^\prime$ in diameter so complete coverage was
not possible; small portions of the remnant toward the south and west
were not imaged. In addition, we point out that the imaging quality
varies across the S3 field of view.  For this observation of \g292\
the aimpoint and its surrounding region of $1^{\prime\prime}$ imaging
(50\% encircled energy radius) is on the eastern half of the remnant.
At the extreme northern and southern extent of the SNR, image quality
has degraded to $\sim$$1.5^{\prime\prime}$, while on the extreme western
edge it has become $\sim$$2.5^{\prime\prime}$.

\section{Spectral and Spatial Analysis}

In Figure 1 we show images of \g292\ in soft (left panel) and hard
(right panel) X-ray bands.  The soft X-ray image shows a highly
structured and filamentary remnant with features on all scales from
arcminutes to arcseconds. The brightest emission is mostly confined to
an irregular ``belt'' that runs in an east-west direction across the
remnant.  These features tend to be quite narrow
($\lsim$3$^{\prime\prime}$) in the transverse (generally north-south)
direction.  Comparison of separate images made in energy bands
containing the strong O, Ne, and Mg K$\alpha$ lines reveals complex
spectral variations with position. The integrated spectrum of the
remnant (Figure 2) shows that lines from these species, in addition to
Si and S, dominate the X-ray emission.  Iron appears not to be a major
constituent of the X-ray spectrum. There is no significant Fe K-shell
emission: the equivalent width of a narrow line feature with energy
between 6.4 and 6.9 keV in the integrated ACIS-S spectrum is $<$200 eV
(at 95\% confidence). We defer a more detailed quantitative discussion
of the thermal emission from \g292\ to a future study.

In the higher energy band, we spectacularly confirm the previous radio
and X-ray evidence for a plerion in \g292. Embedded within a diffuse
nebula approximately 1$^\prime$ in radius, we find a point-like source
at position $\alpha_{\rm J2000} = 11^{\rm h}24^{\rm m}39.1^{\rm s}$,
$\delta_{\rm J2000} = -59^{\circ}16^{\prime}20.0^{\prime\prime}$,
which we designate CXOU~J112439.1$-$591620 (hereafter the ``pulsar
candidate''). We verified absolute positions in the ACIS image by
identifying two serendipitous X-ray sources with optical stars
($m_R=10$ and $m_R=11.8$) from the USNO-A2.0 catalog. Our resulting
X-ray positions are accurate to $<$1$^{\prime\prime}$.

Figure 3 shows a smoothed 4--8 keV X-ray image centered on the pulsar
candidate. This band was chosen to isolate the hard nebular emission
and represents a compromise between obtaining sufficient signal from
the nebula and avoiding contamination by the generally softer thermal
emission.  The pulsar candidate, located near the peak of the diffuse
emission, has a FWHM extent of 1.3$^{\prime\prime}$, while the
unrelated source nearby (which is at roughly the same off-axis angle)
has an extent of 1.1$^{\prime\prime}$, which we take to infer that the
pulsar candidate has an extended component (perhaps a terminal wind
boundary).  (Neither source shows an obvious optical counterpart in
the digitized POSS.)  As the insert to Fig.~3 shows, the X-ray
emission from the immediate vicinity of the pulsar candidate is rather
complex with a symmetric, extended component (out to a radius of
$\sim$1.5$^{\prime\prime}$) embedded in a ridge of emission that is
aligned nearly east-west. On larger scales, the pulsar candidate is
centered on an arcminute-long ridge of fainter emission that is also
oriented in roughly the east-west direction.  The pulsar candidate is
located about 0.9$^\prime$ from our eyeball estimate of the geometric
center of the remnant, which is indicated by the plus sign on Fig.~3.

Within a radius of 2.5 pixels (1.23$^{\prime\prime}$) centered on the
position of the pulsar candidate we obtain a total of 3326 ACIS-S
events for a count rate of $0.0773 \pm 0.0014$ s$^{-1}$. Background,
both instrumental and from the rest of \g292\ itself, was sufficiently
small in this aperture to be neglected.  An absorbed power-law
provided an acceptable fit to the \chandra\ data.  Numerical
values from the fits are given in Table 1 and the spectrum is plotted
in Fig.~2.  The unabsorbed fluxes of the pulsar candidate are
$6.9\times 10^{-13}$ ergs cm$^{-2}$ s$^{-1}$ (0.2--4 keV band) and
$4.8\times 10^{-13}$ ergs cm$^{-2}$ s$^{-1}$ (2--8 kev band).

Detection of pulsed emission would clearly confirm the pulsar
hypothesis.  Unfortunately the low time resolution implicit in the
ACIS Timed Exposure mode makes detection of rapid pulsations
impossible.  We searched unsuccessfully for pulsed emission from the
vicinity of the pulsar candidate in the \rosat\ HRI and PSPC data.  In
neither data set was the candidate detected as a resolved source, due
to the poorer angular resolution of these instruments, the strong soft
thermal X-ray emission, and the low statistical signal. In addition
there have been no reports of pulsed X-ray emission from \g292\ from
either ASCA or RXTE.

The spectrum of the diffuse nebula is significantly contaminated by
the remnant's soft thermal emission within annular apertures even
rather close to the pulsar candidate. For example, the aperture
extending over 3$^{\prime\prime}$ -- 5$^{\prime\prime}$ unmistakeably
shows soft thermal emission (i.e., prominent lines of O, Ne, and Mg),
which introduces significant uncertainty into the derived spectral
properties of the hard, diffuse nebula. Nevertheless out to a radius
of 20$^{\prime\prime}$ we were able to confirm that the spectrum above
3 keV is a power law with photon index in the range $\alpha_p = 1.7 -
2.0$.  Because of the thermal contamination it was not possible to
determine if the photon index of the nebular emission steepened with
radius.  Such an effect has been seen in other Crab-like SNRs with
\chandra\ (e.g., G21.5$-$0.9; Slane et al.~2000) and is expected for
an extended synchrotron nebula powered by a central pulsar.

The total X-ray flux of the diffuse nebula is an important observable
that we estimated in the following manner.  In the 4--8 keV band the
radial profile of the diffuse nebula centered on the pulsar candidate
fell roughly as an exponential with radius. The profile was summed out
to the radius where the emission was twice the background level, which
yielded a total count rate of 0.145 s$^{-1}$. For comparison the
pulsar candidate count rate in the same energy band is only 0.0074
s$^{-1}$, while at the other extreme, the total background subtracted
rate from \g292\ is 0.2188 s$^{-1}$. Thus the diffuse nebula accounts
for 66\% of the total 4--8 kev flux from the remnant.  We assumed the
nebular spectral parameters to be the same as those of the pulsar
candidate in order to convert the band-limited count rate to fluxes.
The estimated nebular spectrum is shown in Fig.~2 (as the middle
curve).  Our estimates of the unabsorbed fluxes from the nebula are
$1.5\times 10^{-11}$ ergs cm$^{-2}$ s$^{-1}$ (0.2--4 keV band) and
$1\times 10^{-11}$ ergs cm$^{-2}$ s$^{-1}$ (2--8 kev band).  The hard
band flux would increase by only about 10\% if we assume that the nebula's
spectrum is steeper, with $\alpha_p = 2.1$ like the Crab. The increase in
the soft band flux in this case would be much larger, about a factor of 2.

\section{Discussion and Results}

We have presented evidence for a point-like source embedded in a
diffuse nebula, both of which exhibit featureless hard X-ray emission
and are spatially coincident with a flat-spectrum radio source in the
SNR \g292.  These features are the classic signatures of a young
rapidly-rotating pulsar and its associated synchrotron, or pulsar
wind, nebula.  The spin-down energy loss rate of the pulsar can be
estimated using the established correlation between this quantity and
the nebular X-ray luminosity (e.g., Seward and Wang 1988).  From our
flux estimate we obtain an unabsorbed luminosity (0.2--4 keV band) of
$L_X = 4\times 10^{34} D_{4.8 \rm\, kpc}^2$ ergs s$^{-1}$, assuming
the H{\sc i} absorption distance of 4.8 kpc (Saken, Fesen, \& Shull
1992), which implies only a modest spin-down energy loss rate of $\dot
E \sim 7\times 10^{36}$ ergs s$^{-1}$.  Based on this estimate, then
for a braking index $n = 3$, and assuming that the current spin period
$P$ is much longer than the initial value, the remnant age implies $P
\sim 280$ ms and a spin-down rate $\dot P \sim 3 \times
10^{-12}{\rm\ s\ s}^{-1}$. This corresponds to a surface magnetic
field strength $B \sim 10^{13}$~G, which is quite large. This value is
reduced if we relax the constraint that $P \gg P_0$.  We note that
there are considerable uncertainties in this estimate, including the
relationship between $L_x$ and $\dot E$ (see, e.g., Chevalier 2000).
Clearly discovery of the pulsed emission from the new pulsar candidate
is of highest priority.  Observations optimized for fast timing using
the HRC on \chandra\ have been approved for observation in cycle 2 and
will be carried out soon.

The pulsar candidate is not directly at the geometric center of the
remnant, suggesting that it has moved since birth.  Its implied
transverse speed of $\sim$$770 (D/4.8\,{\rm kpc}) (t/1600\,{\rm
yr})^{-1}\,\rm km \,s^{-1}$ is in good agreement, given our
uncertainties, with the mean birth velocity of pulsars of
$450\pm90\,\rm km \,s^{-1}$ (Lyne \& Lorimer 1994).  Concerning the
uncertainties in our velocity estimate we note that it is likely that
the blast wave has expanded more slowly toward the southeast in \g292.
This is where the optical emission is most intense and thus where the
density of the ambient interstellar medium should be highest. If true
this would shift the true center of the remnant toward the southeast
and thereby reduce the implied transverse velocity of the pulsar
candidate.  We note that there is no evidence in the X-ray band for a
bow shock preceding the pulsar candidate in this, or any, direction.

To summarize, the \chandra\ data reveal the effects of a central
pulsar on three spatial scales in \g292: an unresolved source (the
pulsar candidate itself); a compact resolved region of X-ray emission
surrounding the point source and an extended pulsar wind nebula.  The
compact emission near the pulsar (see insert to Fig.~3) shows a
symmetric component, $\sim$1.5$^{\prime\prime}$ (0.035 pc) in size, as
well as an elongated ridge of emission covering roughly
$3^{\prime\prime} \times 10^{\prime\prime}$ ($0.07 \times 0.23$ pc).
Such compact, but resolved, components have been seen by \chandra\
near the central compact objects in other known pulsar-wind nebula,
specifically the Crab (Weisskopf et al.~2000) and G21.5$-$0.9 (Slane
et al.~2000). These structures, on scales of tenths of parsecs, have
been interpreted as arising from the pulsar wind termination shock
(e.g., Rees \& Gunn 1974, Kennel \& Coroniti 1984).  On larger scales
the triangular morphology of the extended pulsar wind nebula in \g292\
bears considerable similarity as well to the overall shape of the Crab
Nebula. The physical sizes are similar as well. The bright portion of
the nebula in \g292\ between the light and dark contours in Fig.~3 is
roughly 1 pc across, which is just about the same size as the long
extent of the so-called X-ray torus in the Crab Nebula.

During its first year of operation \chandra\ has discovered point-like
sources within the two youngest Galactic oxygen-rich supernova
remnants (Cas A and G292.0+1.8) that are likely the compact remnants
of these SN explosions.  The compact object in Cas A remains
mysterious (e.g., Pavlov et al.~2000), while the one in \g292\ appears
to have characteristics that are consistent with a young
rapidly-rotating pulsar like the Crab pulsar.  Interestingly enough
Cas A and \g292 also differ in the properties of their X-ray emitting
ejecta. Cas A is dominated by Si- and Fe-rich ejecta that were likely
produced by explosive O- and Si-burning (Hughes et al.~2000).  In
\g292, on the other hand, we have been unable to find evidence for
either of these particular types of nucleosynthetic yields, while the
most common ejecta features are rich in O, Ne, and Mg.  Further study
should reveal whether these differences in the X-ray emitting ejecta
are due to an evolutionary or age effect or whether they signal more
fundamental differences in the stellar progenitors.  Such studies
using \chandra\ will offer us a new window on the connection between
compact objects and the SN progenitors that form them.

\acknowledgments

This research has made use of the SIMBAD Astronomical Database
maintained by the Centre de Donn\'ees astronomiques de Strasbourg.
Financial support was provided by NASA contract NAS8-38252 (PSU), NASA
contract NAS8-39073 (SAO), and \chandra\ grant GO0-1035X to Rutgers.

\newpage

\begin{deluxetable}{lc}
\tablecaption{Power-law Spectral Model Fits for Pulsar Candidate}
\tablewidth{4.75truein}
\tablehead{
\colhead{Parameter} & \colhead{Value and Uncertainty (1 $\sigma$)}
}
\startdata
$N_{\rm H}$ (H atoms cm$^{-2}$) & $3.17\pm0.15 \times 10^{21}$ \nl
$\alpha_{\rm P}$                & $1.72\pm0.05$ \nl
$F_{\rm E}(1\, \rm keV)$ (photon s$^{-1}$ cm$^{-2}$ keV$^{-1}$)
  & $1.44\pm 0.07 \times 10^{-4}$ \nl
$\chi^2$/d.o.f           &  116.1/102 \nl

\enddata
\end{deluxetable}

\newpage

\figcaption[fig1.ps] {\chandra\ ACIS-S X-ray images of \g292\
in soft (left: 0.6--2 keV) and hard (right: 2--7 keV) X-ray bands. The
hard band clearly shows a diffuse, plerionic nebula containing a
central point-like source. The grayscale display for the hard band has been
adjusted so that the diffuse emission can be seen clearly, which has
``burned in'' the point-like source causing it to appear extended.
Note that the point source can also be seen in the soft band image,
where it appears essentially unresolved.
\label{fig1}}

\figcaption[fig2.ps] {ACIS-S X-ray spectrum of the entire \g292\
supernova remnant (data points at top), the estimated power-law model
for the diffuse nebula (middle curve), and the hard point source
(``pulsar candidate'') with best-fit power-law model (data points and
curve at bottom). Above 4 keV the diffuse nebula accounts for 66\% of
the total flux from \g292. The pulsar candidate spectrum extracted
from circular region 2.5 pixels (1.23$^{\prime\prime}$) in radius was
rebinned to 25 counts per channel before fitting to ensure appropriate
Gaussian errors for $\chi^2$ fitting.
\label{fig2}}

\figcaption[fig3.ps] {\chandra\ ACIS-S 4--8 keV band X-ray image of
the pulsar wind nebula in \g292.  The diffuse emission was adaptively
smoothed to an approximate signal-to-noise ratio of 10, while the
unresolved sources were smoothed using a single gaussian with $\sigma
= 0.5^{\prime\prime}$.  The contour levels are plotted at values of
0.0173, 0.0254, 0.0372, 0.0547, 0.0802, 0.118, 0.173, 0.619, 2.22 cts
s$^{-1}$ arcmin$^{-2}$.  The insert shows the raw data in the
immediate vicinity of the pulsar candidate in unblocked pixels (i.e.,
0.492$^{\prime\prime}$ $\times$ 0.492$^{\prime\prime}$).  The
brightest pixel contains 45 detected events. Note the roughly
east-west extension of the central X-ray emission on both large and
small spatial scales.  The plus sign in the main image panel marks the
approximate center of the soft thermal emission of the remnant.
\label{fig3}}

\clearpage

\plotfiddle{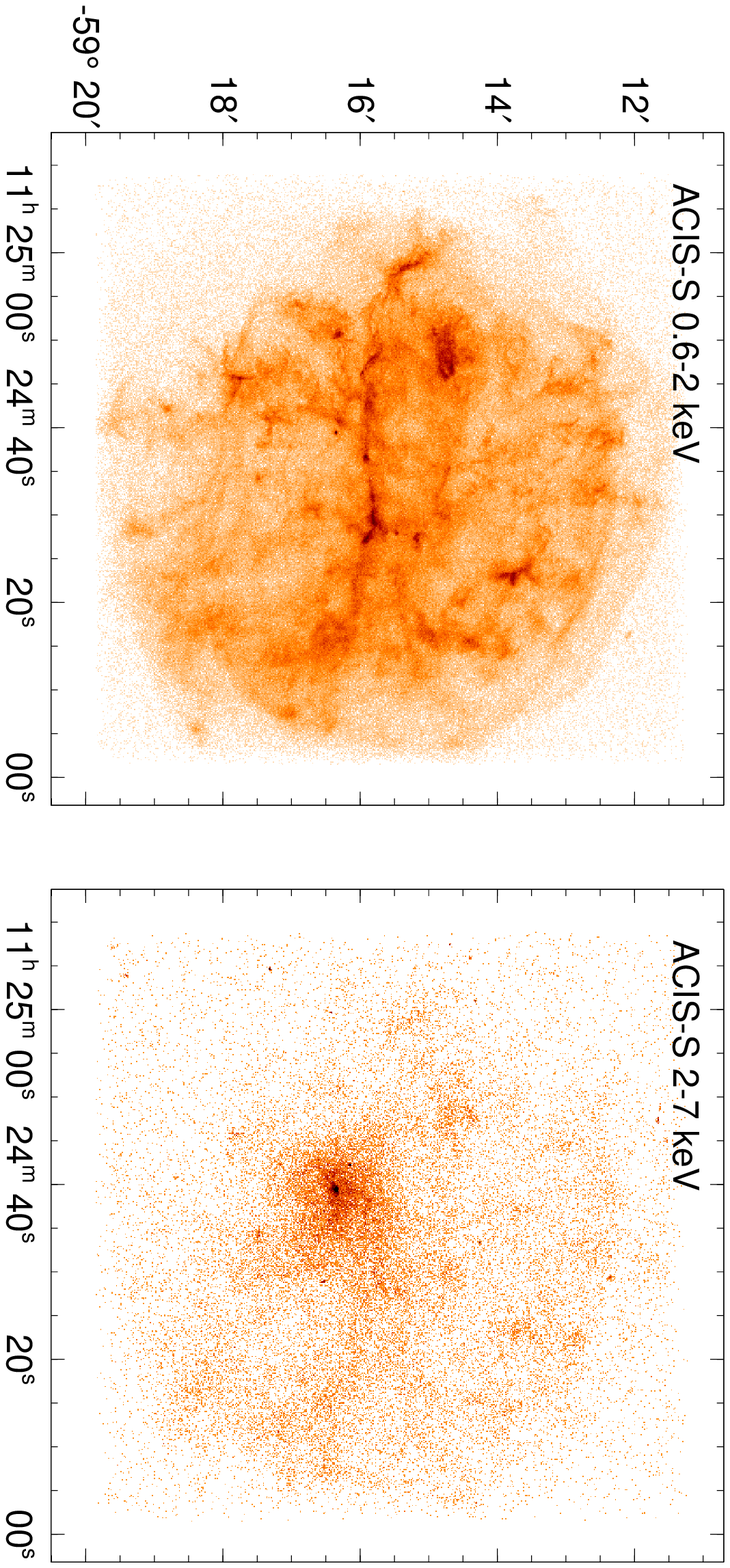}{7in}{0}{100}{100}{-250}{-200}

\clearpage

\plotfiddle{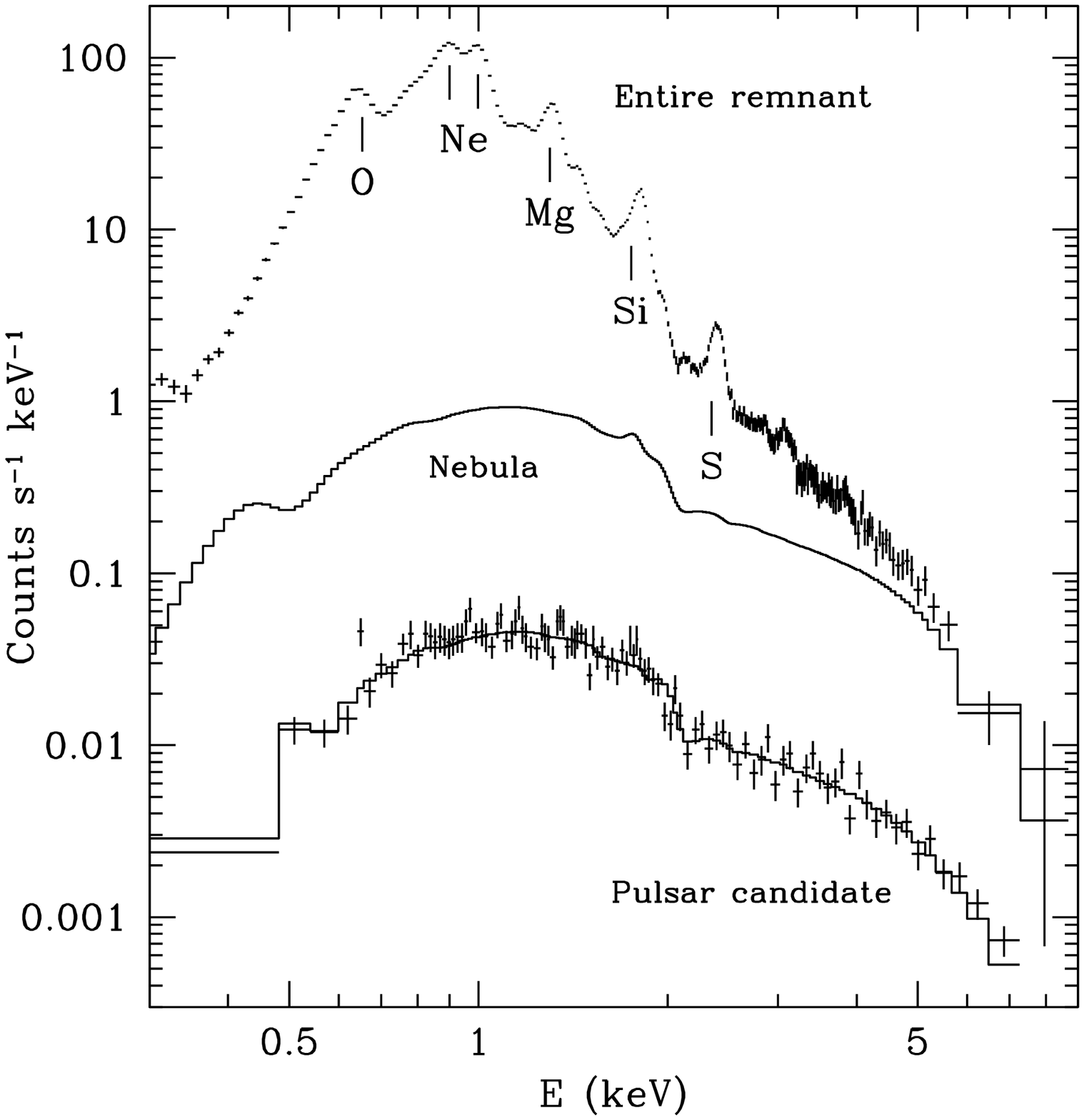}{5in}{0}{75}{75}{-240}{-200}

\clearpage
 
\plotfiddle{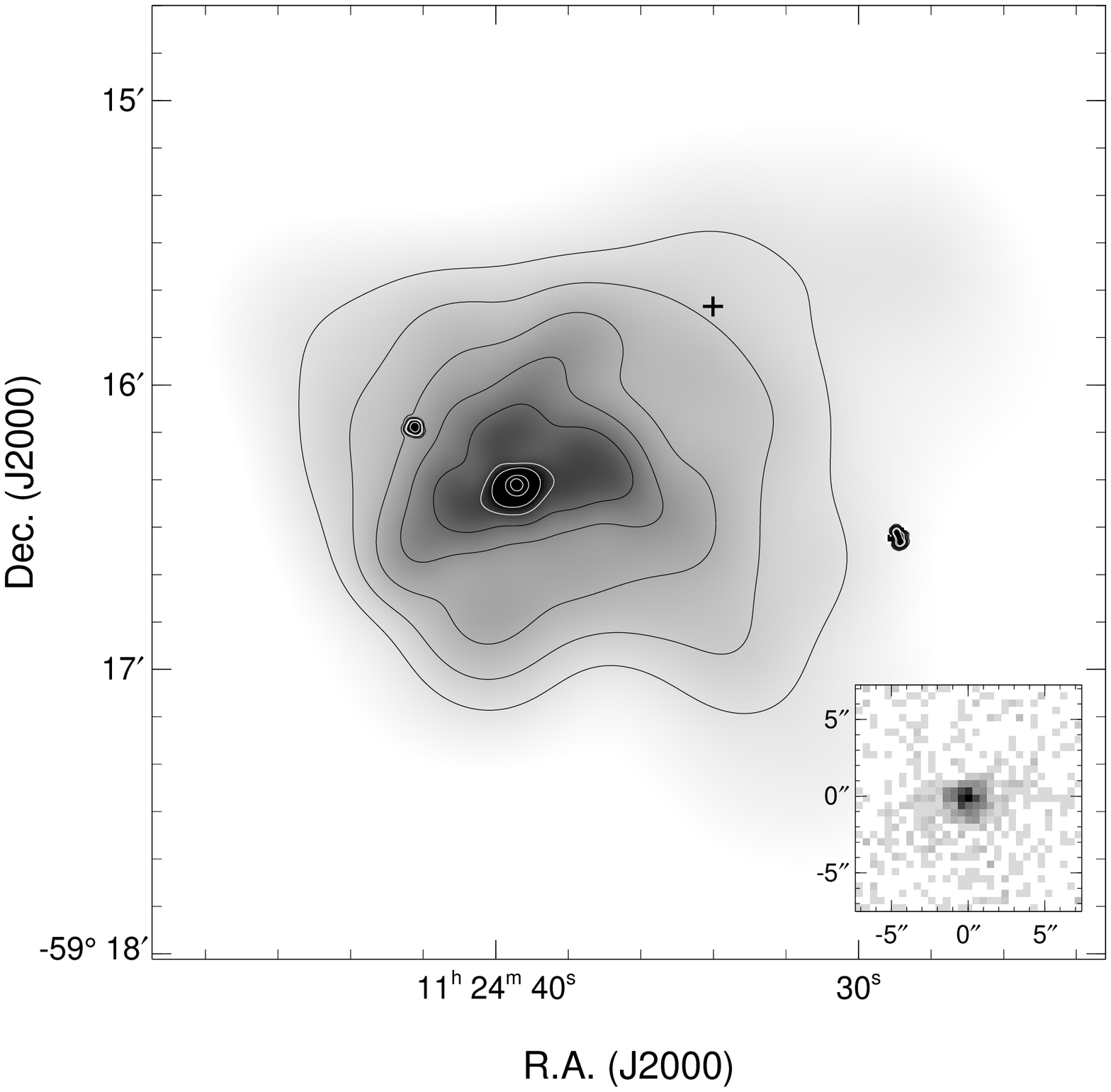}{5in}{0}{75}{75}{-240}{-200}

\end{document}